# Novel two-dimensional photocatalyst $SnN_3$ for overall water splitting with enhanced visible-light absorption

Shengyao Wu[†], Yanqing Shen[†,*], Xu Gao, Yanyan Ma and Zhongxiang Zhou

Department of Physics, Harbin Institute of Technology, Harbin 150001, China

† Contributed equally to this work

* Corresponding author, shenyanqing2004@163.com

**Abstract:** We propose a novel excellent two-dimensional photocatalyst $SnN_3$ monolayer using first-principles calculations. The stability of $SnN_3$ monolayer have been examined via formation energy, phonon spectrum and ab initio molecular dynamics calculations. Large optical absorption capacity plays significant role in the enhancement of photocatalytic splitting of water. The $SnN_3$ monolayer have ultra-high optical absorption capacity in visible region, which is as three and four times as that of $SnP_3$ and $MoS_2$ monolayer, respectively. Available potential and appropriate band positions indicating the ability of overall water splitting even in a wide strain range. Electronic properties of $SnN_3$ monolayer can also be engineered effectively via the external strain, such as the conversion from in-direct band gap to direct band gap. The applied electric field splits the energy levels due to Stark effect, resulting in states accumulation and smaller gap width.

## 1. Introduction

Since Fujishima and Honda discovered photocatalytic splitting of water on semiconductor electrodes in 1972 [1], a clean and effective method to produce hydrogen comes into our sight. From then on, over 100 catalyzers, including $TiO_2$, $WO_3$, $Bi_2WO_6$, ZnO, $Bi_2O_3$ and CdS have been found to play an important role in hydrogen production [2-6]. However, the long distance that photo-generated electrons and holes have to migrate before reaching the solid & water interface leads to the high carrier recombination rate [7]. Therefore, two-dimensional (2D) materials are favored for the tiny migrate distance, and the giant specific surface area also further enhance the photocatalytic efficiency [7].

Molybdenum disulfide ($MoS_2$) is a kind of layered 2D compound of graphene-like, which belongs to hexagonal crystal [8-10]. It is the favored catalyst for overall water splitting because of its appropriate chemical potential and band positions [11, 12]. However, the poor light absorption capacity in infrared and visible spectrum limit its sunlight utilization, and the recombination of photo-generated electrons and holes resulting in low photocatalytic efficiency. Hu et.al replaced S atom in layered $MoS_2$ into Se atom to obtain $MoSe_2$ [12]. Based on first-principle calculation, they found $MoS_2$ and $MoSe_2$ monolayers have band gaps of 1.95 eV and 1.40 eV, respectively. And the electronic properties of $MoX_2$ (X=S, Se) can be effectively modified by applied strain.

$AP_3$ (A=In, Ge, Sn) are the novel graphene-like 2D materials with great electrical properties and optical absorption capacities. According to the reports [13, 14], Miao *et al.* theoretically predicted a single layer of indium triphosphide ($InP_3$). They showed that this is a 2D semiconductor with high stability, suitable band gap of 1.14 eV, and high electron mobility of $2\times10^3$ $cmV^{-1}s^{-1}$. The bulk $GeP_3$ was experimentally synthesized in 1970s, but its monolayer and bilayer nanostructures just have been studied recently [15]. Different from the bulk configuration, the monolayer $GeP_3$ has a band gap of 0.56 eV at the PBE level, and it can be engineered by the biaxial stain [16, 17]. Besides, the $GeP_3$ monolayer also has better carrier mobility ($10^3$ $cmV^{-1}s^{-1}$) than $MoS_2$ [18], but much smaller than monolayer and bilayer $SnP_3$. Layered $SnP_3$ is a new type 2D material predicted from $GeP_3$ with ultrahigh carrier mobility ($1.15\times10^4$ $cmV^{-1}s^{-1}$) and great light absorption capacity [18-22]. Compared with other materials like $GeS_2$ (0.52 $Jm^{-2}$) [23, 24], NaSnP (0.81 $Jm^{-2}$) [25], $GeP_3$ (0.91 $Jm^{-2}$) [26], $GaN_2$ (1.09 $Jm^{-2}$) [27], $InP_3$ (1.32 $Jm^{-2}$) [13] and graphene (0.36 $Jm^{-2}$) [28, 29], the calculated cleavage energy of 1L and 2L $SnP_3$ are 0.57 $Jm^{-2}$ and 0.38 $Jm^{-2}$ [20]. Therefore it is viable to exfoliate from the bulk. Moreover, P. L. Gong *et al.* found that the monolayer $SnP_3$ possesses great chemical stability.

Our $SnN_3$ monolayer is a theoretical prediction based on density functional theory (DFT) computation. The calculated formation energy from $SnP_3$ to $SnN_3$ is -220 meV/atom, comparable with Pt dope in $MoS_2$ monolayer (-230meV/atom) [30]. Phonon spetrum and ab initio molecular dynamics (AIMD) are also proceed in this work, their results shows that our structure has high stability. To better understand the electronic properties of $SnN_3$ monolayer, band natures are computed with both Perdew-Burke-Ernzerhof (PBE) functional [31] and Heyd-Scuseria-Ernzerhof 2006 (HSE06) functional [32, 50-51]. Different functionals give the similar band structures, appropriate band gap straddle the potential of hydrogen and oxygen evolutions, indicating the ability of overall splitting of water. As a photocatalyst of water splitting, the optical absorption is worth characterizing, especially in the visible region. Therefore we calculate the real and imaginary parts of dielectric constants then using Kramer-Kronig relations [33] to obtain the absorption coefficient of $SnN_3$ monolayer. The result is fantastic that ultra-high absorption appears in visible spectrum, which is three times as much as $SnP_3$ monolayer and four times as much as $MoS_2$ monolayer, respectively. Moreover, band structure of $SnN_3$ monolayer can be effectively modified via biaxial strain and external electric field can significantly reduce the band gap which attributes to the Stark effect. Overall, our contribution provides a thorough prediction on the optoelectronic and photocatalytic performances of $SnN_3$ monolayer, which indicates the significant application in the fields of novel nano-electronic devices and photocatalysts for water splitting reactions.

## 2. Computation method

All first-principles calculations are performed using plane-wave based density functional theory (DFT) [34] integrated in the CASTEP [35, 48] software package. In order to proceed geometric optimization and system energy computation, the complex interaction of electron-ion is simplified by the norm conserving pseudopotential (NCP) [36].

To get more accurate electrical and optical properties, we optimize both scale of lattice and atom positions with the local density approximations (LDA) of Perdew and Zunger parameterization of the numerical results of Ceperley and Alder (CA-PZ) functional [37, 49], the energy cut-off is set to 770 eV for the plane-wave basis set. As convergence standards for the structural relaxation, the remnant stress and force is under 0.02 GPa/atom and 0.01 eV/Å, respectively. Meanwhile, atom displacement should be smaller than $5\times10^{-4}$ Å, and the energy change is converged to within $5\times10^{-6}$ eV/atom.

In the self-consistent field (SCF) calculations of density of states (DOS) and other properties, we use GGA/PBE functional with ultra-soft pseudopotential (USP) [38], choosing an energy cut-off of 500 eV for the plane-wave basis set and a k-mesh of 5×5×1 to sample the Brillouin zone (BZ). Considering that GGA/PBE functional usually underestimates the band gap, HSE06 functional with NCP takes over the band structure calculation to get the accurate gap width, 770 eV plane wave energy cut-off and 3×3×1 k-point grid is enough to guarantee the correct results. Because the optical properties calculated by HSE06 functional are inaccurate while they are strongly related to the gap width. So we use the optical data calculated by GGA/PBE and scissor the under estimated part of the band gap then proceed the absorption calculation. As a measure to proceed 2D material calculation, our cell is built with a 17 Å vacuum slab along Z direction to avoid the interlayer periodic interactions.

## 3. Result and discussion

### 3.1 Structure and stability

Based on the previous achievements [18, 20-22], our $SnN_3$ start from the $SnP_3$ monolayer (R-3m space group), replacing the P atoms into N atoms then proceeding geometry optimization to minimize the remnant force and stress. Figure 1 (a) exhibits the completely optimized lattice from top, and the top view and the side views of $SnN_3$ 2×2 supercell are shown in Figure 1 (b) - (d). After the optimization, the structure of 2D $SnN_3$ belongs to the symmetry group P-3M1. The structural parameters of $SnN_3$ monolayer are listed in Table 1, compared with $SnP_3$, it has a smaller lattice parameter of 5.25 Å (7.14 Å for $SnP_3$), which caused by the closer combination of N-N bonds than P-P bonds. It is obvious that all the N atoms settle on the same plane, and Sn atoms distribute symmetrically beside N-plane. In order to evaluate possibility of synthesis, we calculate the formation energy $E_{form}$ and binding energy $E_{bind}$, their expressions are:

$$E_{form}=E_{SnN3}-E_{SnP3}-nE_N+nE_P \tag{1}$$

$$E_{bind}=E_{SnN3}-aE_{Sn-atom}-bE_{N-atom} \tag{2}$$

Where $E_{SnN3}$ and $E_{SnP3}$ are the total energy of SnN$_3$ and SnP$_3$ monolayer, n represents the number of N atom in SnN$_3$. $E_N$ and $E_P$ refer to the normalization energy of nitrogen and phosphorus, respectively. $E_{Sn-atom}$ and $E_{N-atom}$ represent the single point energy of Sn atom and N atom, the quantity of Sn atom and N atom in SnN$_3$ unit cell are shown by a and b.

As results, the formation energy from SnP$_3$ to SnN$_3$ reaches -220 meV/atom, comparable with Pt dope in MoS$_2$ monolayer (-230 meV per atom) [30] and in a low position of the formation energy alignment [7]. So the SnN$_3$ system has great structural stabllity. Moreover, the binding energy $E_b$ of SnN$_3$ monolayer is -6.36 eV/atom.

Phonon spectrum is usually used to evaluate the material stability. We proceed phonon dispersion calculation with finite displacement method integrated in CASTEP package, and the cut-off radius of supercell is set to 8.5 Å (the volume of supercell is 12 time as the unit cell). As shown in Figure 2 (a), the minimum phonon frequency appears at G point while it is still positive, that no imaginary frequency phonon is found at any wave vector, indicating our SnN$_3$ structure is dynamically stable. Further, we compare the highest frequency phonon mode of monolayer SnN$_3$ with that of SnP$_3$, MoS$_2$ and silicone monolayers. The highest frequency phonon mode of monolayer SnN$_3$ reaches 1386 cm$^{-1}$, much higher than that of monolayer silicone (580 cm$^{-1}$) [52], monolayer SnP$_3$ (500 cm$^{-1}$) [20] and monolayer MoS$_2$ (473 cm$^{-1}$) [54], it can be explained by the strong ionic bonds between Sn and N atoms. We also proceed the thermodynamic simulation in its 3×3×1 supercell with Dmol3 package. Considering that pressure is not a significant factor, we choose the constant volume and constant temperature ensemble (NVT), and the initial temperature is set to 300 K. The total simulation time is 2800 fs with a time step of 1 fs. Results are shown in Figure 2 (b), the energy vibrates in the first 250 steps, then maintains within 10$^{-3}$ eV/atom during the rest of the steps (much smaller than SnP$_3$ monolayer [20]), indicating that the structure reaches dynamic equilibrium after 250 fs. The 3×3×1 SnN$_3$ cell maintains its hexagon structure and the N-N honeycomb ring even unchanged undergoing this 2800 steps simulation. That is, the material is dynamic stable at 300 K.

Electrostatic potential indicates the minimum energy that electron must absorb when it escapes from a material, which shows relationship with chemical stability [18, 20-22]. Figure 3 (b) presents the work function of $SnN_3$ monolayer, the calculated Fermi level locates at -5.02 eV, lower than $GeP_3$ monolayer (-4.89 eV), $SnP_3$ monolayer (-4.88 eV) and phosphorus monolayer (-4.25 eV), indicating this material is chemically stable.

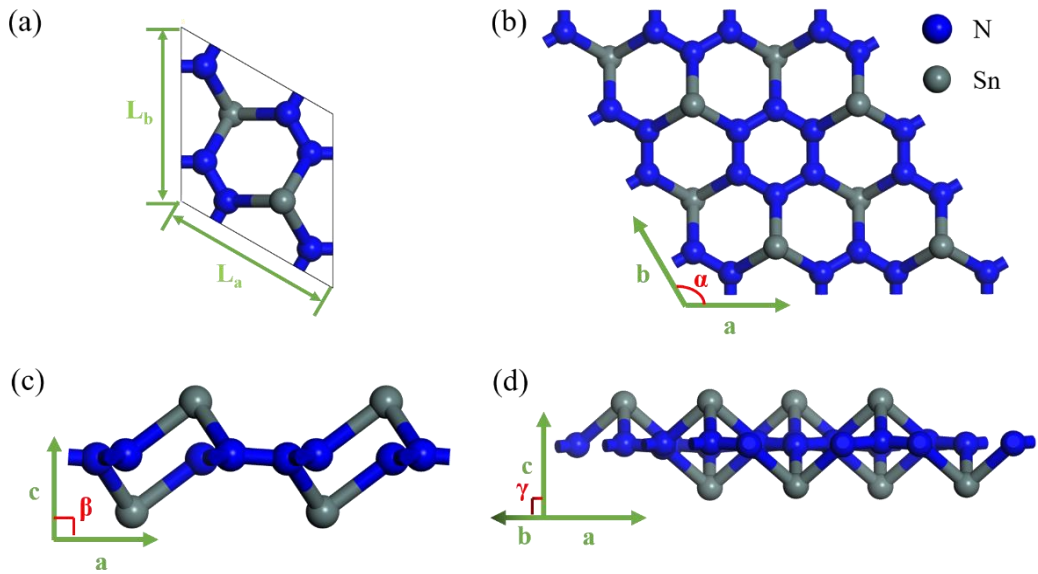

Figure 1. (a) The top view of unit cell. Top view (b) and side views (c), (d) of $SnN_3$ monolayer.

Table 1. Lattice parameters $L_a$ and $L_b$, lattice angles α, β and γ, bond lengths $L_{N-N}$ and $L_{N-Sn}$, bond angles ∠$_{N-N-N}$ and ∠$_{N-Sn-N}$ of $SnN_3$ monolayer.

| $L_a$ (Å) | $L_b$ (Å) | $L_{N-N}$ (Å) | $L_{N-Sn}$ (Å) | α (deg) | β (deg) | γ (deg) | ∠$_{N-N-N}$ (deg) | ∠$_{N-Sn-N}$ (deg) |
|---|---|---|---|---|---|---|---|---|
| 5.25 | 5.25 | 1.37 | 2.09 | 90.00 | 90.00 | 120.00 | 119.46 | 87.17 |

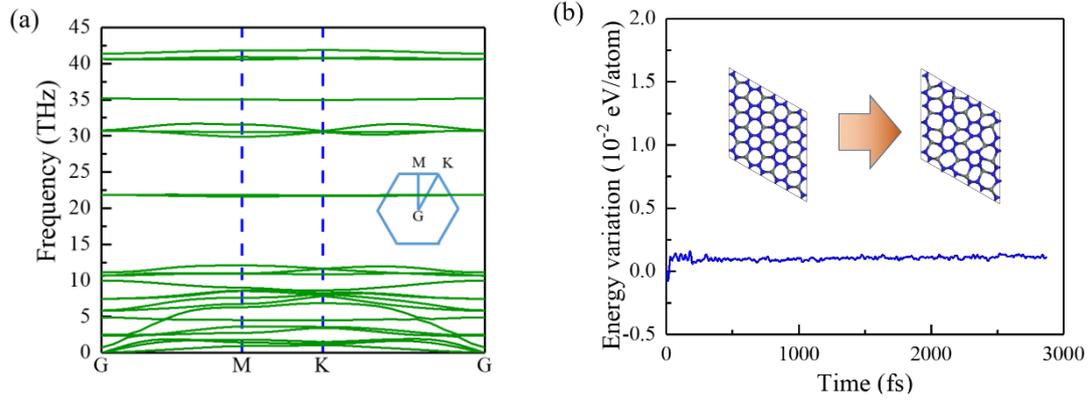

Figure 2. (a) Phonon spectrum of monolayer $SnN_3$. (b) Energy variations in AIMD at room temperature (300K).

## 3.2 Properties

Figure 3 (a) exhibits calculated band structures of $SnN_3$ monolayer with GGA/PBE functional (black solid lines) and HSE06 functional (purple solid lines), the blue dashed line marks the Fermi level. The spin-orbital-coupling (SOC) of the relatively heavy atom Sn is not considered in the calculation because it shows little effect on band structure [20]. Different methods give similar band structures, forbidden band is wide near G point and becomes narrower between M and K point. When scrutinizing the band structures data, the result calculated by PBE functional shows primary direct transition at K with a band gap of 1.410 eV, and we find another direct transition at M point which required energy is just 0.11 eV higher than the primary transition. The HSE06 functional presents a primary in-direct transition which valence band maximum (VBM) at M point while conduction band minimum (CBM) at K point while a direct transition at M point with only 0.1 eV higher than this in-direct transition. Considering that GGA/PBE functional usually underestimates width of band gaps, we report the HSE06 result of 1.965 eV as precise width of the band gap.

Water splitting is a clean and effective way of oxygen and hydrogen production. However, to proceed water redox, some restrict conditions should be satisfied [7]. One is the band gap not only should exceed the free energy of water splitting of 1.23 eV but also should be smaller than 3 eV to guarantee the reaction sufficiency. The other is the band edges must straddle oxygen evolution reaction (OER) level and hydrogen evolution reaction (HER) level. That is, the electrostatic potential of CBM is higher than -4.44 eV while that of VBM shall lower than -5.67 eV. The Fermi level of SnN$_3$ monolayer has an appropriate electrostatic potential at -5.02 eV shown in Figure 3 (b), fixing its CBM at -4.04 eV and VBM at -6.00 eV, which is 0.40 eV upper than HER level and 0.34 eV lower than OER level, respectively, exhibiting the ability of overall water splitting. As the previous theoretical prediction, the closer CBM and HER level are (the same to VBM and OER level), the higher electrolytic speed will be [7]. However, the over potential of electrode often limit the efficiency of water reduction and oxidation which have been reported by many experimental studies [39-41]. Thus, the redundant potentials of SnN$_3$ monolayer might guarantee the water splitting reaction sufficiently proceed.

Optical absorbance is a significant ability for photocatalytic splitting of water, we use optical absorption coefficient α to measure this ability, which is expressed by:

$$\alpha(\omega) = \sqrt{2}\omega \left[\sqrt{\varepsilon_1^2(\omega) + \varepsilon_2^2(\omega)} - \varepsilon_1(\omega)\right]^{\frac{1}{2}} \quad (3)$$

Where ε$_1$ and ε$_2$ are the real and imaginary parts of the complex dielectric constant, they are both functions of the frequency $\omega$. Because of the anisotropy, ε$_1$ and ε$_2$ both are second-order tensors, if we assume $\alpha$ and $\beta$ are the two directions in Cartesian coordinate, the matrix element $\varepsilon_{\alpha\beta}^2$ is expressed by:

$$\varepsilon_{\alpha\beta}^2(\omega) = \frac{4\pi^2 e^2}{\Omega} \lim_{q \to 0} \frac{1}{q^2} \delta(E_{ck} - E_{vk} - \hbar\omega) \langle u_{ck+e_{\alpha}q} | u_{vk} \rangle \langle u_{ck+e_{\beta}q} | u_{vk} \rangle^* \quad (4)$$

In the function, $\Omega$ is unit cell's volume, three-dimensional vector $\vec{q}$ denotes the Bloch vector of the incident wave and scalar $q$ is one of its component in Cartesian coordinate. The eigenstates of vector $k$ in VB and CB are represented by $u_{ck}$ and $u_{vk}$, the energy of them are $E_{ck}$ and $E_{vk}$, respectively. Then we can figure out the $\varepsilon_{\alpha\beta}^1$ via Kramer-Kronig relations [33]:

$$\varepsilon_{\alpha\beta}^1(\omega) = 1 + \frac{2}{\pi} P \int_0^\infty \frac{\varepsilon_{\alpha\beta}^2(\omega')\omega'}{\omega'^2 - \omega^2 + i\eta} d\omega' \quad (5)$$

Where $\eta$ is an infinitesimal number and $P$ refers to principle value.

Figure 4 (b) presents the in-plane light absorption coefficients of SnP$_3$ monolayer, SnN$_3$ monolayer and MoS$_2$ monolayer as functions of photon energy from 0 eV to 10 eV. Comparing with $SnN_3$, $SnP_3$ and $MoS_2$ shows better light absorption in ultra-violet part. However, almost all the solar energy is stored in infra-red and visible region, and considering that the energy difference between HER level and OER level, the photon energy used for single-photon photocatalysis should be higher than 1.23 eV. Therefore, visible light absorption plays the key roll in photocatalytic splitting of water. As shown in Figure 4 (c), due to its small band gap of 0.7 eV [18, 20-22], $SnP_3$ monolayer have a large absorption peak at infrared and red light range, but get poor absorption in the rest visible spectrum. As for $MoS_2$, large band gap and small density of states (DOS) near CBM and VBM leads to a shortage absorbance in infrared and visible region. The absorption coefficient of $SnN_3$ monolayer is 3 and 4 times as that of $SnP_3$ and $MoS_2$ monolayer in visible region, respectively. Absorption capacity shows great relationship with DOS, Figure 4 (a) exhibits the partial density of states (PDOS) of $SnN_3$ monolayer, the 2 states peaks closest to VBM and CBM (marked with green backgrounds) strongly contribute to the light absorption peak at 2.1 eV and 2.5 eV. The abundant bound N-p states electrons absorb photon energy and transit to free states which consist of Sn-p and P-p guarantees the high value of light absorbance. Moreover, the absorption valley highlighted by green arrow in Figure 4 (c) consistent well with the lack of states at -1.5 eV and 1.5 eV in DOS (marked with red dot lines in Figure 4 (a)).

Figure 4 (d) exhibits the electron density difference (EDD) of $SnN_3$ monolayer, the yellow parts represent the electron depletion and the regions in light blue refer to charges accumulation. Because of the different electronegativity between Sn and N atoms, the electrons originally belong to Sn atoms are attracted by the N-N honeycomb rings. It can be obviously observed that the VBM is mainly dominated by the hybrid nitrogen states corresponding to the floriform yellow regions between the N atoms in the EDD. Last but not the least, electron localization function (ELF) is also proceed to study the bonding condition. As it shown in Figure 4. (e), the electrons between Sn and N atoms present weak localized character that the Sn-N bonds are ionic bonds. It proved by the great difference in electronegativity between Sn and N.

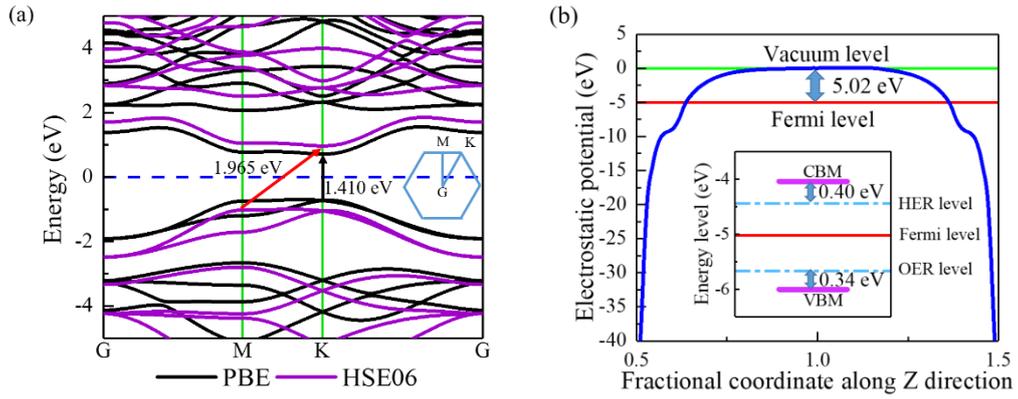

Figure 3. (a) Calculated band structure of SnN$_3$ monolayer with PBE and HSE06 functional. (b) Electrostatic potential along Z direction, green and red solid lines refer to vacuum and Fermi level, respectively. The small figure in the center marks the CBM, VBM and Fermi level of SnN3 monolayer, hydrogen evolution reaction (HER) level and oxygen evolution reaction (OER) level are also presented in this figure.

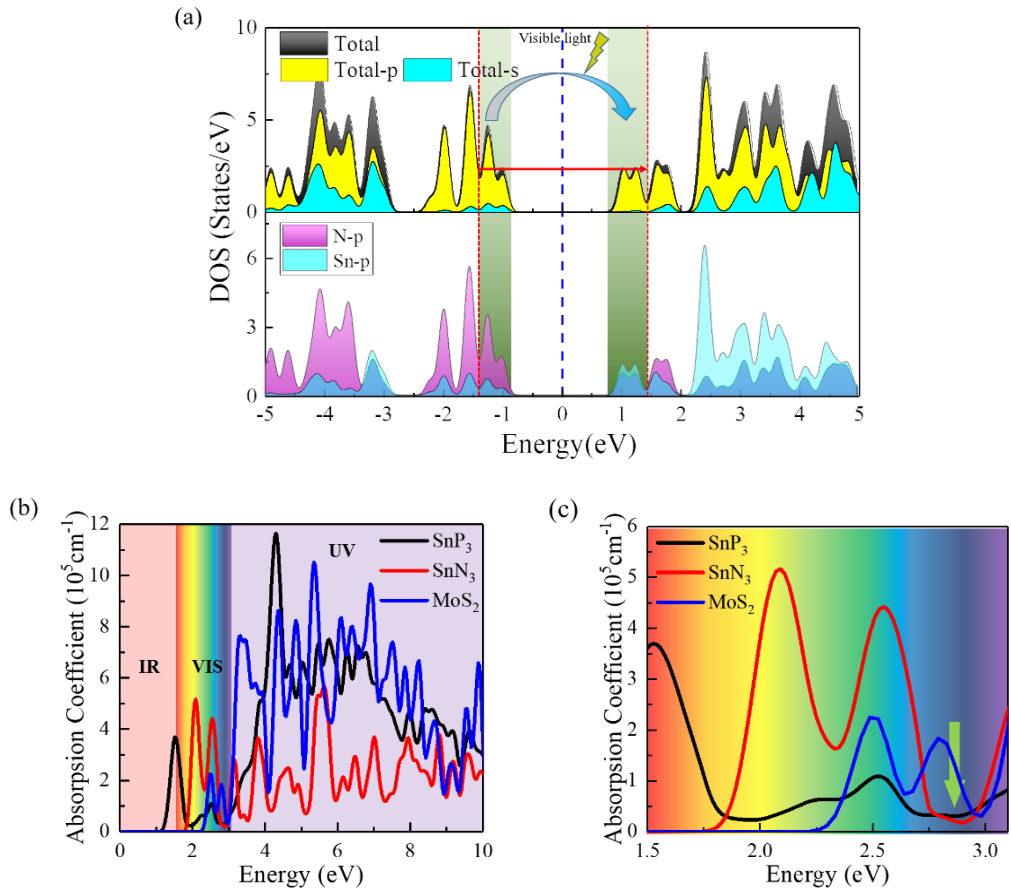

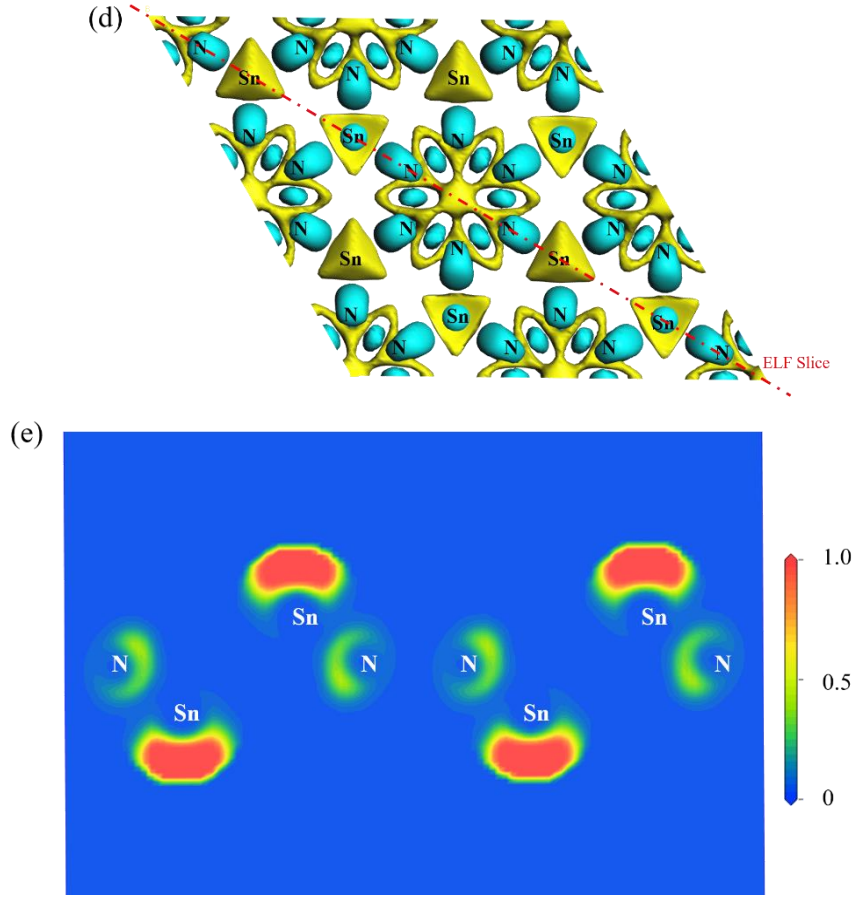

Figure 4. (a) Partial density of states (PDOS) of $SnN_3$ monolayer in its 1×1×1 cell (6N and 2Sn). (b) Optical absorption in the photon energy range of 0 eV – 10 eV and (c) visible light absorbance of $SnN_3$ monolayer. (d) Electron density difference (EDD) of monolayer $SnN_3$ 2×2 supercell, the isosurface is set to 0.08 e/Å³. (e) The electron localization function (ELF) of electrons between N atoms and Sn atoms, the direction of the slice is figure out by the dot dash line in (d).

## 3.3 Under Strain

Many properties of mechanical, electronic and optical are influenced tremendously by strain engineering [42]. Compressive or tensile strains are also helpful in solving the ineluctable lattice parameter mismatch problems in the construction of heterostructures. The controllable band structures brought by strain engineering broaden the applications of 2D materials. Figure 5. (a), (b) show the band structure of $SnN_3$ monolayer changes with the tensile and compressive biaxial strain along both zigzag and armchair directions, (c) presents the band gap widths in different strain regions. Tensile strain in 0% ~ +8% range causes a little overall drop of the CBM, which narrower the band gap.

We also find that the second excited state at G point becomes significant lower as the strain increases (marked with colored arrows on the right side), when strain reach +8%, the first and second conduction bands (CBs) above Fermi level are intersected, if further increased the tensile strain, the CBM becomes lower near G point, changing the structure of CBM. This trend also causes a little reduction of CBM at M point when the strain is +12%, resulting in the direct band gap at M point.

Compressive strain from 0% ~ -6% slightly increases the band gap while band gap width decrease with the further enlargement of the strain. When deformation finally comes to -12%, the band gap is 1.405 eV at HSE06 level. Moreover, the VBM at M point descend under biaxial compressive strain, leading a transformation from in-direct band gap to direct band gap. The advantage of direct band gap is that electrons can transit between CBM and VBM at K point without momentum change, which means the phonons do not need to be involved in the whole process and all the energy can be released via photons, indicating high luminous efficacy. Which illustrates that $SnN_3$ monolayer have great potential of applications in light emitting devices brought by strain engineering.

Lattice deformation can largely influence the electron potential. As shown in Figure 5. (c), with the increasing of the compressive biaxial strain, chemical potential firstly goes down to -5.20 eV (+4%), then rises up to -4.75 eV (-12%). Interestingly, the band gap also varies with the external strain, so the CBM of $SnN_3$ monolayer under different compressive strains can always maintain around -4.00 eV. Tensile strain rises the Fermi level of $SnN_3$ monolayer, obviously, resulting in the elevation of CBM and VBM. Considering that water splitting needs electrons potential under -5.67 eV and holes potential above -4.44 eV, $SnN_3$ is suitable for decomposing water in a wide strain range from -10% to +8%.

Figure 5. (d) presents the light absorption conditions of $SnN_3$ monolayer under compressive strains (in warm colors) and tensile strains (in cool colors). The material under tensile strains maintain the same absorption structure as the original one, only the peaks slightly move left. However, compressive strains significantly change the absorption curve, splitting the peak at 2.6 eV into two and shifting them to 2.7 eV and

2.9 eV, which enhances the absorption capacity of purple light (2.8 eV to 3.0 eV). The absorption edges move with the deformation of the lattice, in agreement with the band gap variations.

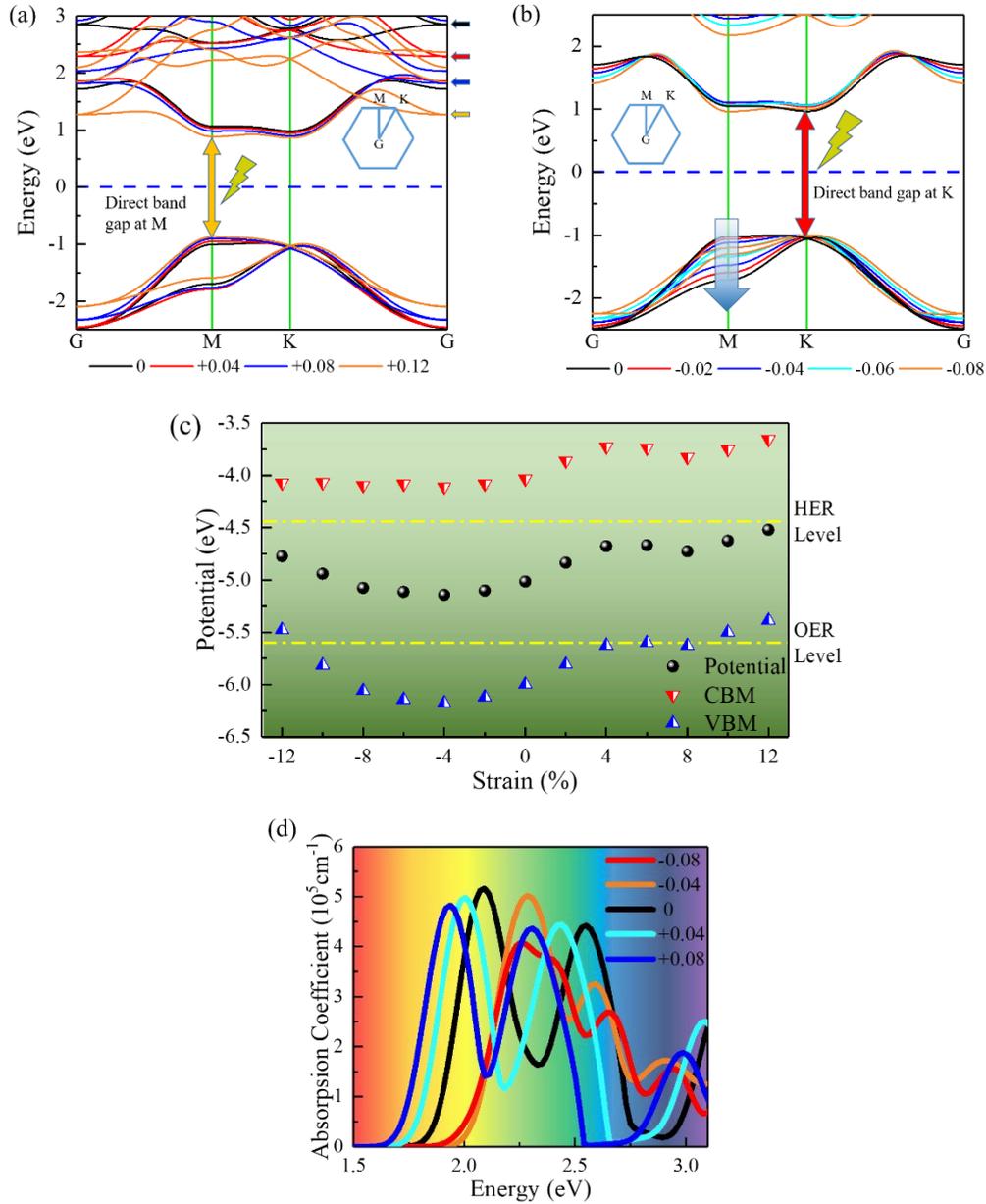

Figure 5. Band structures under tensile biaxial strain from 0% to +12% (a) and compressive biaxial strain from -12% to 0% (b) at HSE06 level. (c) Shows the band gap width calculated by PBE and HSE06 functionals under different strain situations. The visible light absorption variation presents in (d).

## 3.4 Under electric field

Triggering a lot of research activities in material science, the external applied electric field can effectively influence the electronic properties of materials, including

2D materials. The achievements obtained on new materials are looked forward to play significant roles in applications including memory elements, quantum computing and low-power consumption devices [43-45].

In order to find the conversion voltage from semiconductor to conductor, we apply the external perpendicular electric field on $SnN_3$ monolayer along Z direction. The increment of the electric field is set to 0.2 V/Å as initial. For the position of abrupt change of band gap, the increment is minimized to 0.05 V/Å. And we calculated the band gap under all voltage with both PBE functional and HSE06 functional. Figure 6. (a) shows the band gap as a function of the in-plane perpendicular electric field. As the electric field gradually increases, a gentle decline happened to the gap width. This character can be clearly explained by the DOS shown in Figure 6. (c), states number grows with the electric field, and the newly generated states distribute asymmetrically besides the original state (more states with lower energies and less with the higher). The increase in the dipole moment (ionic bonds) upon electrical excitation, which called first-order Stark effect [46], resulting in the symmetrically growth of states. And the asymmetry of new states comes from the second-order Stark effect, it caused by the excess of the polarizability, a nonlinear effect at ultra-high voltage (E>$10^7$V/m [47]). The second-order Stark effect splits the original states into new states with lower energies, explaining the movement of states near CBM to Fermi level. To prove the splitting of states, we integral the states from -3 eV to +3 eV. Total quantities of states are shown in Figure 6. (d), the states in VB are marked yellow while CB states are green. When the electric field intensity increases, the states in CB growth in a significant trend, however, the VB states affected little by the electric field, indicating that the valence electrons are more stable than free electrons.

Figure 6 (b) exhibits the band structure of $SnN_3$ monolayer under different electric fields. It can obviously observed that the electric field strongly affects CBs, but influence little on the structure of valence bands (VBs). The new bands split by the external electric field mainly appear around +2 eV to +3 eV when the electric field is 0.4 V/Å. And if the field intensity comes to 0.8 V/Å, the original CBM divides into 2 bands, one of them becomes the new CBM which edge locates at G point, indicating an in-direct character of band gap. Moreover, the further increase of electric field moves CBM to lower energy level and finally cross the Fermi level. That is, complete the conversion from semiconductor to conductor, the conversion limit is 1.2 V/Å.

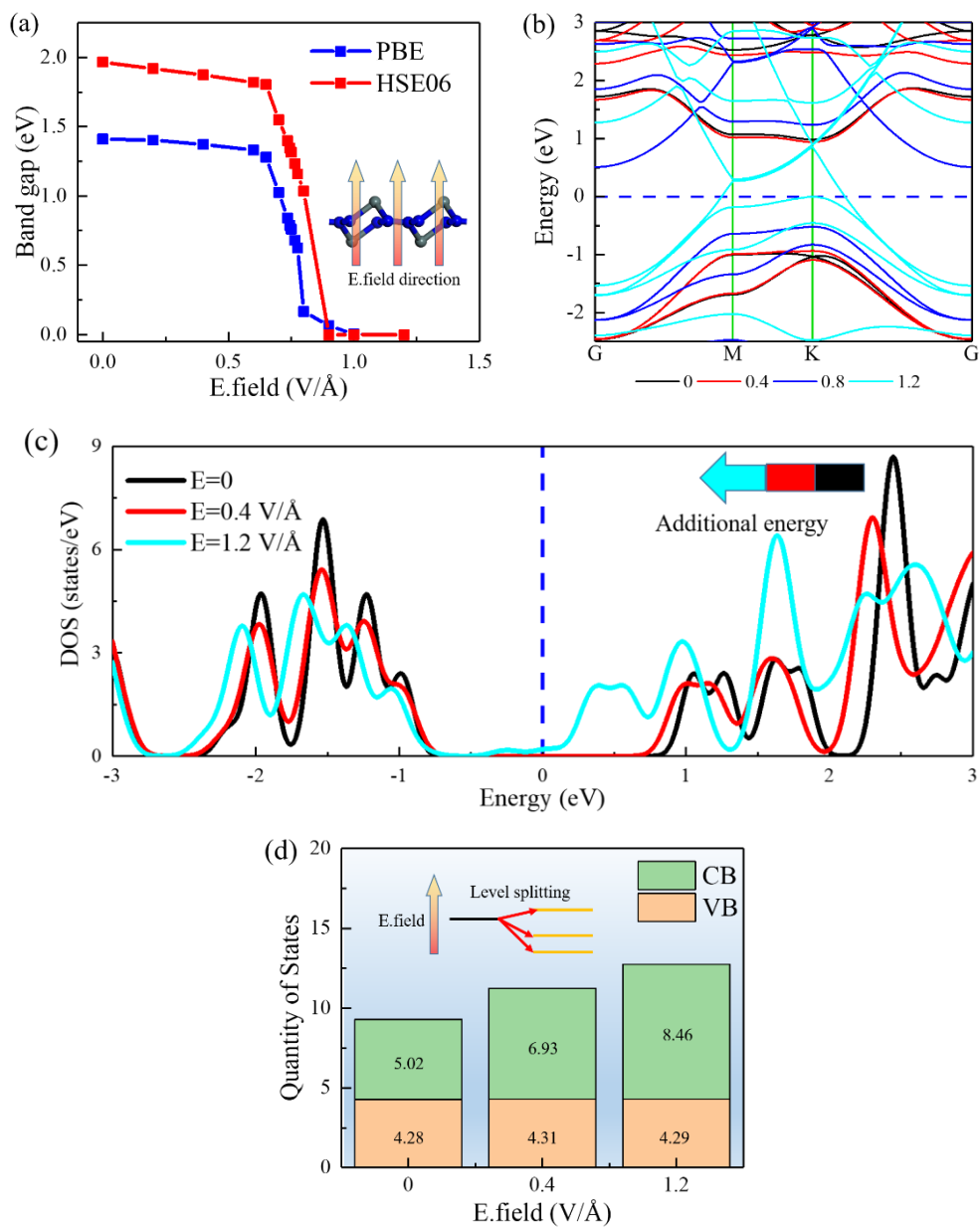

Figure 6. (a) The band gap of SnN$_3$ monolayer as a function of applied electric field at PBE and HSE06 level. (b) Band splitting under different external electric fields. (c) The movements of DOS caused by different external electric fields. (d) States quantities in the energy range of -3 eV ~ +3 eV under different electric fields.

## 4. Conclusion

Herein, we proceed the first-principle calculation screening and identified novel 2D materials SnN$_3$ monolayer. Firstly, we evaluate the possibility of synthesis using formation energy (-220 meV/atom) and binding energy (-6.36 eV/atom). Secondly, the phonon spectrum and AIMD simulation also show that the structure of SnN$_3$ monolayer have great dynamics stability. As electronic properties, the electrostatic potential is -5.02 eV, which makes the appropriate band gap of 1.965 eV across the HER and OER levels, indicating the application potential in overall splitting of water. The band structure can be effectively modified via the external strain, in particular the conversion from in-direct band gap to direct band gap. Also, the biaxial tensile strain can significantly elevate the chemical potential while the potential is reduced via compressive strain. In a wide strain range, the material maintains the ability of overall water splitting. We use absorption coefficient figured out by K-K relations to evaluate the light absorption capacity of SnN$_3$ monolayer. The material exhibits great light absorbance, in particular in visible spectrum, which is as three and four times as that of in SnP$_3$ and MoS$_2$ monolayers, respectively. Last but not the least, the Stark effect caused by the external applied electric field are also been studied. Conversion voltage from semiconductor to conductor of SnN$_3$ monolayer is about 1.2 V/Å. These results based on first-principle calculation predict the electronic, optical and strain properties of SnN$_3$ monolayer, describing the wide application prospects in semiconductor field, especially in photocatalytic water splitting.

5. Acknowledgement

The research is supported by National Natural Science Foundation of China (No.11204053 and No.11074059) and the China Postdoctoral Science Foundation (No. 2013M531028).